\documentclass[a4paper, amsfonts, amssymb, amsmath, reprint, showkeys, nofootinbib, twoside]{revtex4-2}
\usepackage[english]{babel}
\usepackage[utf8]{inputenc}
\usepackage[colorinlistoftodos, color=green!40, prependcaption]{todonotes}
\usepackage[pdftex, pdftitle={Article}, pdfauthor={Author}]{hyperref} 
\usepackage{empheq,etoolbox}
\patchcmd{\subequations}
  {\theparentequation\alph{equation}}
  {\theparentequation.\alph{equation}}
  {}{}
\begin{document}
\title{Harmonic fractal transformation for modeling complex neuronal effects: from bursting and noise shaping to waveform sensitivity and noise-induced subthreshold spiking}

\author{Mariia Sorokina}
    \affiliation{90 Navigation street, Birmingham, UK, B5 4AA \\ k4939b@gmail.com}

\date{\today} 

\begin{abstract}
We propose  the first fractal frequency mapping, which in a simple form enables to replicate complex  neuronal effects. Unlike the conventional  filters, which  suppress or amplify the input spectral components according to the filter weights,   the transformation  excites novel  components  by a fractal recomposition of the input spectra resulting in a formation of spikes at  resonant frequencies that are optimal for sampling.  
This enables high sensitivity detection, robustness to noise and noise-induced signal amplification.  The proposed model illustrates that a neuronal functionality can be viewed as  a linear summation of spectrum over nonlinearly transformed frequency domain.
\end{abstract}

\maketitle
\section{Introduction}
One of the simplest neuronal models - the Schmitt trigger \cite{Schmitt} - emulates a threshold excitation via a temporal transfer function. To capture complexity of nonlinear multi-scale temporal effects (e.g. bursting \cite{bursting}, high-frequency block \cite{HFB} or subthreshold spiking \cite{noise}) the differential equations or multi-variable mappings defined in temporal domain are used (among the simplest see  the FitzHugh-Nagumo model \cite{FHN1,FHN2} or Chialvo  mapping \cite{Chialvo}).

Recent research reveals novel dynamical effects, for example:  the waveform selectivity in neuronal amplification processes \cite{waveform} or the effect of noise shaping observed in neural networks \cite{eckhorn} and in a single FitzHugh-Nagumo neuron \cite{HFT}. Furthermore, it was shown that the noise shaping effect plays a role in the subthreshold spiking \cite{ANS}. While the Schmitt trigger enables a  \textit{noise suppression} due to the noise averaging, physically tractable models replicating a \textit{noise shaping} mechanism and associated effects in neuronal systems are lacking. 

Here we illustrate  that it is possible to recreate the aforementioned complex effects characteristic to neuronal processing using the  harmonic fractal transformation (HFT) \cite{HFT} as a frequency mapping.   The HFT is a self-similar infinitely repeated rotate-and-scale pattern. It exhibits a new type of semi-analogue quantization: defined in frequency domain it results in a novel intertwined  amplitude-frequency modulation similar to the complex patterns arising in neuronal spiking \cite{HFT}.  The HFT incorporates the frequency mixing effect \cite{FM} by exciting novel frequency components at the signal harmonic nearest to the neuronal frequency. This results in reshaping of the input signal and noise modes into the characteristic neuronal pulseshapes modulated at the characteriestic neuronal frequencies, consequently, increasing the spiking rate. Thus,  a spike amplification and waveform sensitivity is enabled and enhanced by the noise due to the nonlinear properties of neural processing. The HFT reveals a role of fractal frequency transformations in neuronal dynamics and enables a novel and simpler approach to modelling neuronal systems.

\begin{figure*}[!ht]
\centering
\includegraphics[width=\linewidth]{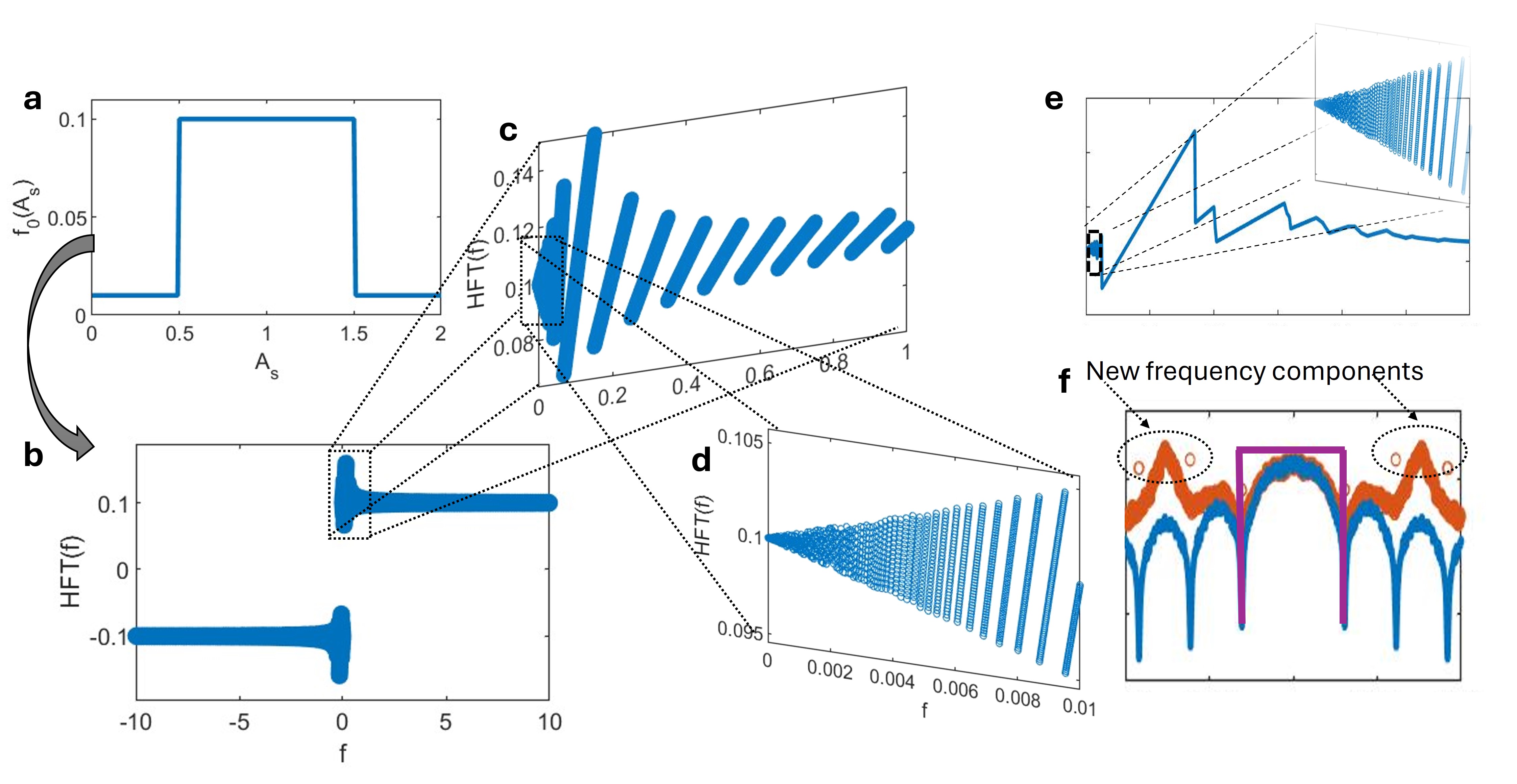}
\caption{\textbf{Frequency transfer function.} 
\textbf{a} Spiking frequency $f_0$ (analogous to pulse-frequency modulation) as a function of stimulus amplitude $A_s$. For simplicity a step-wise function is used:  the spike frequency tend to the characteristic neuronal frequency (here $f^*=0.1$) when $A_s$ is within-the-threshold region, outside which  the output follows the stimulus frequency (here $f_s=0.01$).  \textbf{b} The Harmonic fractal transformation: the input frequency $f$ is semi-quantized to the chosen eigen-frequency $f^*$ in a semi-analogue manner via the fractal property of the transformation, see enlargements in \textbf{c,d} demonstrating an infinite rotate-and-scale pattern. \textbf{e} A  spiking frequency/bandwidth transformation observed in the FitzHugh-Nagumo (FHN) model is schematically plotted here for comparison. \textbf{e} The HFT excites new spectral components around the eigen-frequency $f^*$.}
\label{F1}
\end{figure*}

\begin{figure*}[!ht]
\centering
\includegraphics[width=\linewidth]{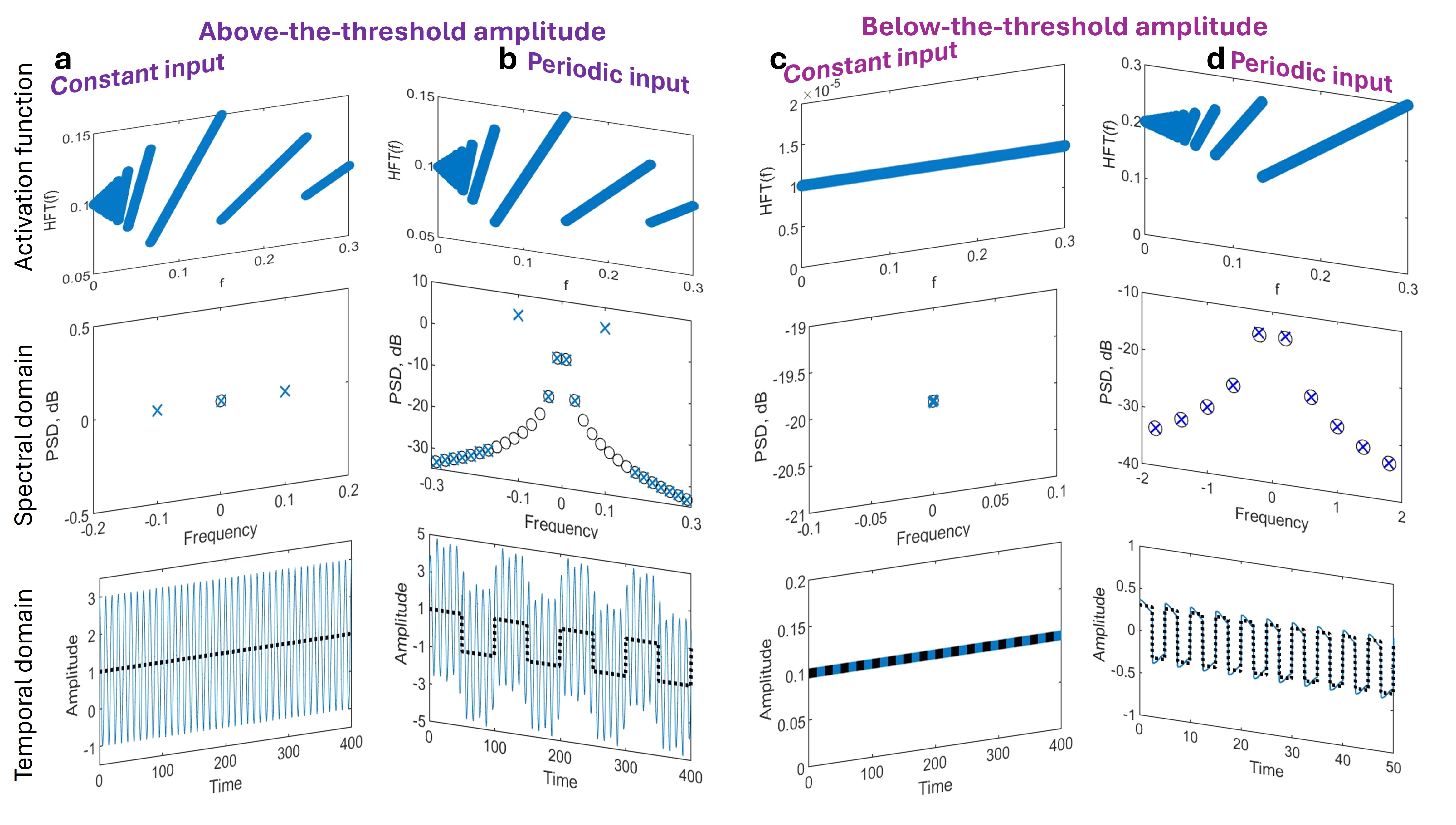}
\caption{\textbf{The role of spectrum in spiking.} 
The HFT (upper panels)  governed by two parameters  $f_0$ and $\hat{f}$  chosen according to the stimulus may (\textbf{a,b}) or may not (\textbf{c,d}) excite the additional spectral components in the output PSD (middle panels). This corresponds to spiking or not spiking in the temporal domain (lower panels). Compare input and output shown in black and blue. 
 At above-the-threshold amplitudes the FHN induces: \textbf{a} regular spiking  for a constant stimulus (here $I_{ext}=1$) and \textbf{b} bursting for slow periodic stimulus (here $I_{ext}=\Pi(f_s=0.01)$). \textbf{c,d} For below-the-threshold stimulus: the FHN preserves the stimulus frequency characteristics: \textbf{c} constant (here $I_{ext}=0.1$) or \textbf{d} periodic (here $I_{ext}=0.3\Pi(f_s=0.2)$). }
\label{F2}
\end{figure*}

\begin{figure*}[!ht]
\centering
\includegraphics[width=\linewidth]{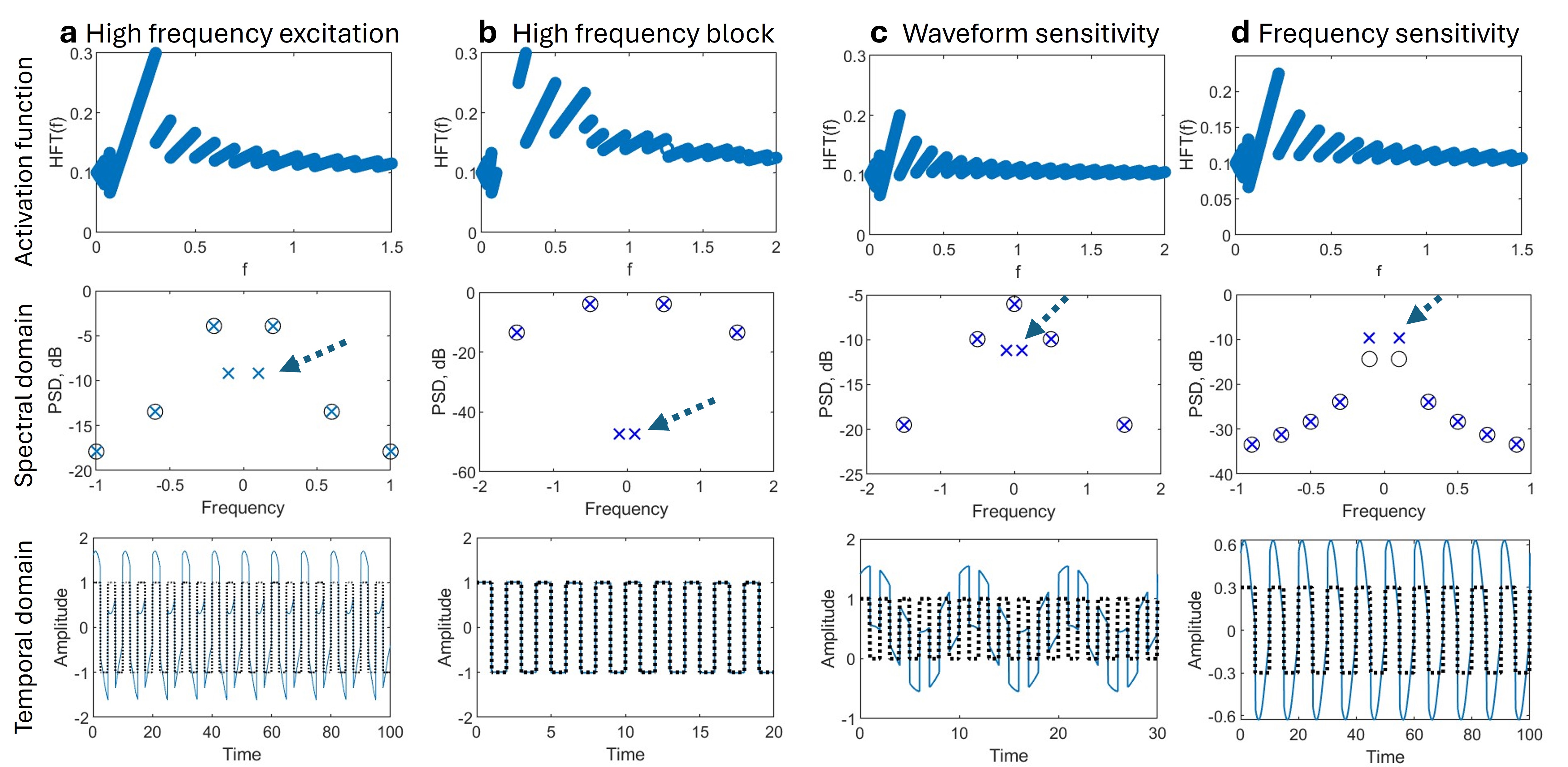}
\caption{
\textbf{Waveform and frequency sensitivity.} 
\textbf{a} For high  frequency stimulus (here $I_{ext}=\Pi(f_s)$ with $f_s=2f^*$) the HFT exhibits a regular pattern excitating a harmonic closest to  $f_0=f^*$ (highlighted by arrows), resulting in a regular spiking. \textbf{b} Higher frequency stimulus (here $f_s=5f^*$) leads to an irregular pattern in the HFT due to the difference in the HFT parameters: $f_0$ and $\hat{f}$ subduing an amplification of the excited frequency components (see arrows), hence, a 
high frequency excitation block occurs. \textbf{c} Changing the waveform (here $I_{ext}=0.5(\Pi(f_s)+1)$ with $f_s=5f^*$) may overcome the excitation block as the HFT (now having $f_0=\hat{f}$) recovers a regular repeated pattern exciting the spectral components and, thus,  spiking. \textbf{d} When the input frequency is close to the eigen-frequency the  HFT exhibits a regular  pattern enabling the spectral components excitation and producing  spiking even at small amplitudes (compare to the same waveform for higher frequency in Fig. 2d).
}
\label{F3}
\end{figure*}

\begin{figure*}[!ht]
\centering
\includegraphics[width=\linewidth]{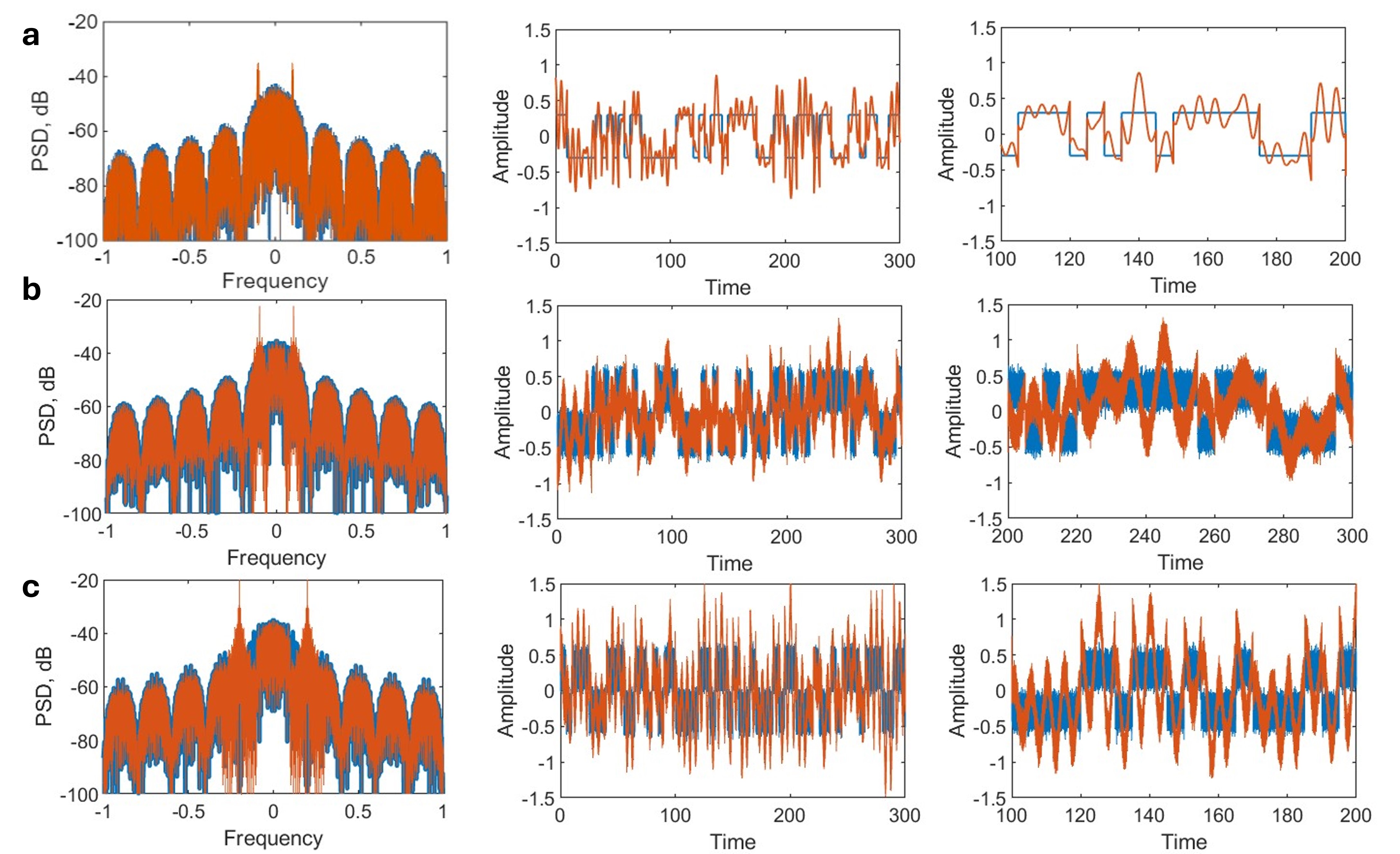}
\caption{
\textbf{Amplification via noise shaping.} 
\textbf{a} Below-the-threshold input (blue) - OOK with $A_s=0.3$ and symbol frequency $f_s=0.2$ after the HFT (here $f_0=0.1$) (red) is spectrally  reshaped resulting in small reshaping in temporal domain (also highlighted in the enlargment). 
\textbf{b} Adding noise to the input increases the peaks at the PSD of the HFT output  and amplifiers the spiking. \textbf{c} Adjusting the HFT to the symbol frequency ($f_0=f_s=0.2$) ensures higher and more frequent spiking.
}
\label{F4}
\end{figure*}

\begin{figure*}[!ht]
\centering
\includegraphics[width=\linewidth]{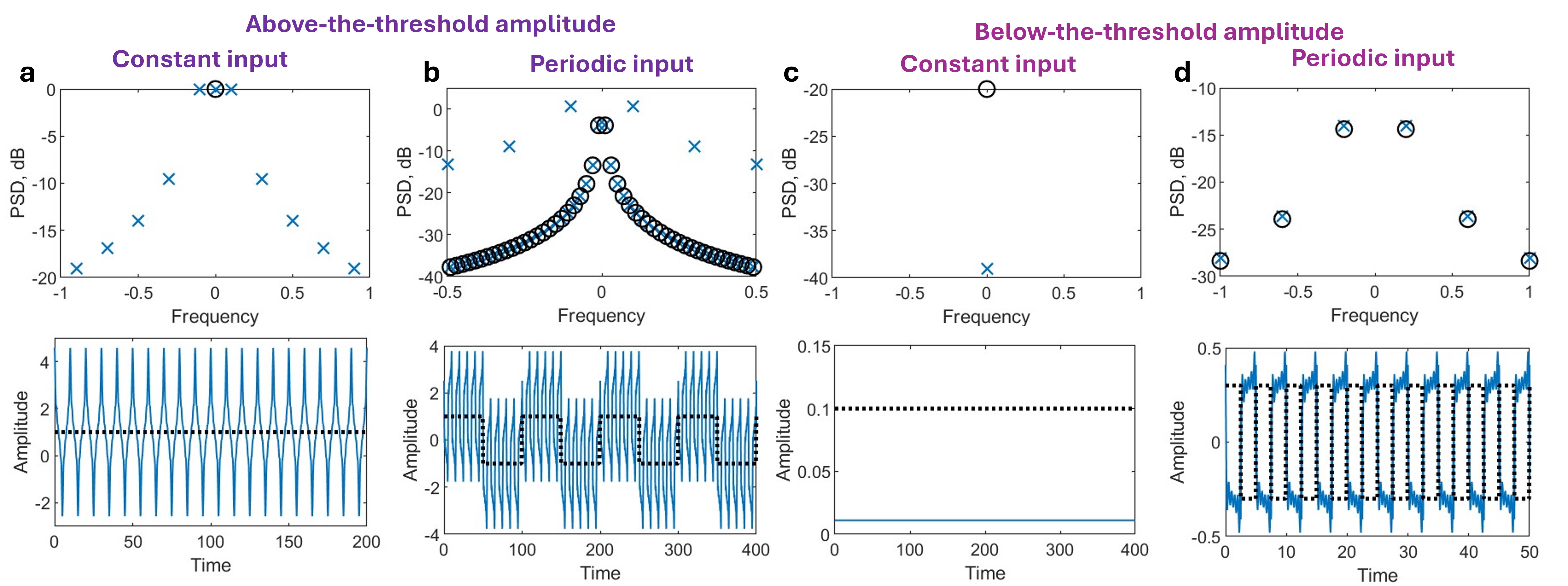}
\caption{
\textbf{Multi-Harmonic Fractal Transformation.} 
Including multiple harmonics enables modeling variety of spiking pulseshapes and vary amplitude response (compare to Fig. 2): \textbf{a,b}. Spiking occurs for above-the-threshold amplitude for input \textbf{a} constant stimulus ($I_{ext}=1$) and \textbf{b} periodic stimulus ($I_{ext}=\Pi(f_s=0.01)$) due to the generation of spectral components at the $nf_s[f_0/f_s]$ harmonics (output in blue and input in black). \textbf{c,d} For below-the-threshold stimulus:  the stimulus frequency characteristics are preserved, here \textbf{c} constant $f_s=0$ (here $I_{ext}=0.1$) or \textbf{d} periodic (here $I_{ext}=0.3\Pi(f_s=0.2)$).}
\label{F5}
\end{figure*}

\section{Harmonic fractal transformation}
The harmonic fractal transformation describes a dependency of the spike frequency on the stimulus frequency \cite{HFT} in the seminal Fitzhugh-Nagumo (FHN) model. This is similar to the pulse-frequency modulation, where the output frequency is  a function of the stimulus amplitude.  Moreover, it was found \cite{HFT} that the FHN generated spikes  have frequencies integer multiple or sub-multiple of input frequency for low or high frequency stimulus, correspondingly, thus, representing a complex frequency-amplitude modulation. This property is highly important for efficient signal sampling and recovery (this is similar but different to phase locking, which is focused on the period of the pulsetrain).  Moreover, it was also noted \cite{HFT} that the FHN spike frequency is the closest integer multiple of the frequency ratio to the characteristic frequency of the FHN neuron when excited by a constant stimulus of the same amplitude.  

These properties allowed us to construct a functional relation in frequency domain. By analysing the spectral properties of the FHN we found that in case of a constant stimulus the generated spike frequency $f_0$ exhibits a step-like dependency on the stimulus amplitude $A_s$ \cite{HFT}. For demonstration we use a simplified step-wise dependency plotted in Fig. 1a:  within-the-threshold region there is an amplitude dependency of spiking frequency,  outside-the-threshold region - the stimulus frequency $f_s$ is preserved (for a constant stimulus: $f_s=0$). We denote the maximum spiking frequency as $f^*$ (here $f^*=0.1$).  Generally, the frequency-amplitude dependence $f_0(A_s)$ can be designed as desired.
The  HFT has a form:
\begin{equation}
\mathrm{HFT}(f)=f\Big[\frac{f_0}{f}\Big]\mathrm{H}(f_0-f)+\frac{f}{\Big[\frac{f}{f_0(1+1/[f/\hat{f}]}\Big]}\mathrm{H}(f-f_0)
\label{E2}
\end{equation}
here $\mathrm{H}$ is the Heaviside function and $[ \ ]$ denotes the nearest integer
function (rounding half up). The parameters governing the HFT are defined by $f_0(A_s)$: $f_0$ is defined by the stimulus amplitude extremum    and $\hat{f}$  by the average  amplitude  $\bar{A}$. Thus, $\mathrm{HFT}(f)$ tends to $ f_0$ and  $ \hat{f}$ for small and  large frequencies, correspondingly. 

As an example consider a rectangular modulated input with an amplitude $A_s=1$ and frequency $f_s=0.01$. From Fig. 1a one obtains: $f_0=\hat{f}=0.1$. The corresponding HFT is plotted in Fig. 1b with infinitely repeated rotate-and-scale patterns highlighted in Figs. 1c,d. An example of the HFT observed in the FHN model \cite{HFT} is  plotted in Fig. 1e  demonstrating a variability  for various frequency parameters. As was shown in \cite{HFT} and schematically illustrated in Fig. 1f, the effect also extends to randomly encoded signals, where $f_s$ represents a symbol frequency. Here a randomly encoded  On-Off Keying (OOK) input with $f_s=f^*/2$ is plotted in blue and the FHN output in red. Note the excited peaks at the eigen-frequency $f^*$. 

For illustration consider a simplified form of the HFT:
\begin{equation}
\mathrm{HFT}^{\mathrm{sim}}(f)=f\Big[\frac{f_0}{f}\Big]\mathrm{H}(f_0-f)+\frac{f}{\Big[\frac{f}{f_0}\Big]}\mathrm{H}(f-f_0)
\label{E22}
\end{equation}
This mapping has similar fractal properties as the full HFT above (while having one parameter $f_0$, instead of two: $f_0$ and $\hat{f}$). Thus, it results in squeezing the input frequency from $(-\infty;\infty)$ into a fractal map, where the output frequency tends to  $f_0$ with  the maximum  deviation $f_0/2$. The unique property of the discovered fractal is that the deviation of the input frequency from $f_0$ is squeezed larger for higher deviation. Or, in other words, the proposed mapping is self-regulated: the higher the deviation from the chosen value of interest $f_0$, the more it is reduced. This property is highly advantageous for sensing as was shown in \cite{HFT} enabling to model the remarkable sensing frequency range observed in biological neurons.
One can use a simplified model $\mathrm{HFT}^{\mathrm{sim}}$ in all of the considered here effects  except the high frequency blocking, where the frequency $\hat{f}$ becomes crucial.
 
In \cite{ANS} we have noted that one can represent the effect of the HFT on signal as an excitation of new spectral components at the frequency-transformed values. These new components are defined as a summation of the initial spectrum at the multi-valued frequencies of the HFT (as an example, $\mathrm{HFT}(f_0)=\mathrm{HFT}(2f_0)$).  For any frequency $q$ the corresponding spectral component $\tilde{A}_q$ is a sum of those input components, whose frequencies  $f$ after the HFT are transformed to  $q$. In other words, it is a sum over spectral components, whose frequencies $f$ satisfy $\mathrm{HFT}(f)=q$:
\begin{equation}
\mathrm{HFT}(\tilde{A}_q) =\sum_{f} \tilde{A}_{\mathrm{HFT}(f)=q}
\label{E3}
\end{equation}
 Note, the number of terms for each $q$ is different, which results in a formation of a peak at the vicinity of the characteristic frequency  $f_0$.
Thus, one can view the HFT as  frequency-dependent summation of the spectral components.  In comparison to the conventional approaches to modeling neural networks, e.g. convolution-based, which use a multiplication of the spectral components, the HFT has  similarity to the Hadamard type-transformations. 
   The HFT convergence is due to the finite nature of signal spectra (while there might be infinite many frequencies composed from $nf_0$ and $f_0/n$ for any integer $n$, the number of non-zero spectral components corresponding to them is limited and convergent). For continuous spectra the summation can be performed over the interval length  $q\in (\mathrm{HFT}(f-\delta f/2),\mathrm{HFT}(f+\delta f/2)]$, where $\delta f$ is the spectral resolution, thus, convergence follows from the underlying Fourier transformation convergence. 
Similarly to the Schmitt trigger and the Delta-Sigma modulation, the HFT enables a noise suppression and has certain quantization properties. Regarding the temporal discontinuities usually associated with modeling the neural network processing (e.g. in digital back propagation), the HFT uses a spectral summation, which may be useful for mitigating the differentiability problem common in spiking neural networks.

There is an interesting analogy to the Wiener-Khinchin-Einstein theorem. Namely, considering not a discrete-time form, but assuming a discrete (fractal) frequency, one can view the described here summation of the signal spectral components at the  discretized frequencies  as an analogue of signal autocorellation in \cite{Einstein} - a statistical quantization  with characteristic neuronal properties. The fractal nature of the transformation enables to achieve an analogue of a characteristic function, similar to autocorrelation. Moreover, this function is resampled and remodulated with a convenient pulseshaping and at convenient frequencies to the neuron, while avoiding aliasing due to harmonic frequencies ratio \cite{HFT}.

\section{Modelling neuronal effects}
\subsection{Frequency transformation as the cause of spiking excitation threshold}
Consider an example of a constant input $I=1$ (see Fig. 2a), from Fig. 1a the corresponding HFT parameters are:  $f_0=f^*=0.1$ (as $A_s=1$)  and $\hat{f}=f^*=0.1$ (as $\bar{A}=1$). The resulting HFT is plotted in the upper panel of Fig. 2a, while the middle panel depicts the spectra. The input spectrum will have a single component at $f=0$, thus $\mathrm{HFT}(f \rightarrow 0)=\pm f_0$. Hence, the output spectrum components are excited at these frequencies as a replication of the central component creating regular oscillations at the output with frequency $f_0$.  

For periodic input let us consider a rectangular waveform with frequency $f_s=0.01$ and unit amplitude (Fig. 2b). The corresponding HFT is similar to the previous example, while having parameters: $f_0=0.1$ and $\hat{f}=f_s=0.01$. 
 One can see here that the non-zero signal components are at frequencies integer multiple of $f_0=0.1$ and, therefore, summed and accumulated at $f_0$  resulting in a peak value at this frequency, while the temporal waveform at the output acquires an $f_0$-oscillation in bursts, while preserving the period of the input.  

For a small constant input (here $I_{ext}=0.1$ in Fig. 2c) the HFT governing frequencies are $f_0=\hat{f}=f_s=0$. The input spectral frequency is transformed as $\lim_{f_s \rightarrow 0}f_s[f_s/f_s]=0$. As  new components are not generated, the input is preserved at the output. Note, that similar to the FHN model, here  spiking  is possible for below the threshold  input  for frequencies close to the neuronal $f_s \simeq f^*$  due to the resonance effect (see frequency sensitivity further below). 
Otherwise, for a small periodic input  (here $A_s=0.3$ and $f_s=0.2$ in Fig. 2d) the HFT governing frequencies are $f_0=\hat{f}=f_s=0.2$ and, while the input is modified at the output, the changes are infinitesimal, i.e. spiking does not occur. Importantly, at the output the period is preserved to the input value, which is also the case in the FHN processing, where at the output the signal is suppressed, yet the period is preserved. 
Thus, within-the-threshold the additional peaks at the eigen-frequency arise and the output is reshaped and amplified, while outside-the-threshold the signal remains largely unaffected and the periodic properties are preserved (see conditions for input suppression further). 

Note,  the origin of spiking is in the excitation of the \textit{additional spectral components} with the HFT acting similar to the finite-impulse response filter but with the threshold-defined parameters shaping the fractal.

\subsection{High frequency block, waveform and frequency sensitivity: an interplay of frequencies}
 We have noted above that due to the fractal nature of the HFT transformation the complex behaviour may arise, which could lead to complex regimes. Let us start with a rectangular wave with the unit amplitude and frequency $f_s=0.2$ (see Fig. 3a). Hence, the HFT parameters: $f_0=f^*$ and $\hat{f}=f_s=2f^*$. The two frequencies do not differ significantly and lead to a regular repeated pattern in the HFT and, consequently, in the output. 

Consider higher frequency $f_s=0.5$  and the same waveform (see Fig. 3b). The HFT will have $f_0=f^*$ and $\hat{f}=f_s=5f^*$. Now the two frequencies differ significantly leading an irregular pattern in the HFT, so that the frequency summation has a number of non-zero components that is too small. This results  in small peaks and  no significant modification of the output waveform, thus, leading to a high frequency block. 

Shifting the waveform up: larger mean value, yet the same maximum amplitude, see Fig. 3c, results in the HFT having $\hat{f}=f^*$ and the same value for  $f_0=f^*$. Now the HFT  results in the larger eigen-frequency components (highlighted by arrows) leading to amplification and reshaping of the temporal waveform (compare with Fig. 3b) . This is similar to  the amplitude-frequency ratio in a frequency block observed in the FHN \cite{HFT}.

Finally, a small input signal modulated at the eigen-frequency (or close value), as plotted in Fig. 3d for $A_s=0.3$ and $f_s=f^*$, excites a spiking output due to the HFT parameters being equal:  $f_0=f_s=f^*=\hat{f}$. This case recreates the frequency sensitivity, i.e. a subthreshold spiking at resonance frequencies in the FHN.

\subsection{Noise shaping and signal amplification}
Let us consider an encoded stimulus. Unlike the periodic cases, which have the discrete harmonics characterized by the \textit{signal frequency} (the inverse of the signal period), a randomly encoded signal has continuous components characterized by the \textit{symbol frequency} (the inverse of symbol period). Here we consider a two-level rectangular modulation - On-Off keying (OOK, i.e. zeros and ones). For an encoded stimulus with $A_s=0.3$ and $f_s=0.2$ (here $f_s$ denotes a symbol frequency) one can compare the input (blue) and the output (red) in Fig. 4a in spectral and temporal domain as well as the corresponding periodic case plotted in Fig. 2d. Here the HFT has the  governing frequencies:  $f_0=0.1$ and $\hat{f}=f_s$, which indices an excitation of the peaks in the spectrum. The finite width of the peaks is due to the continuous nature of the input spectrum and is defined by the signal bandwidth, this is similar to the FHN case \cite{HFT}.    As highlighted in the enlargement of the temporal domain on the right panel in Fig. 4a the consecutive patterns of bits can induce spiking. Coding can be advantageous to induce spiking for below the amplitude threshold stimulus as in continuous spectrum case there are more frequency components, which contribute to the formation of these peaks.  The HFT captures this.

Adding a small amount of noise to the input may further facilitate spiking. As an example we add a white Gaussian noise with variance $0.1$: notice the increased peaks at the eigen-frequencies and the corresponding increase in the spikes amplitude in Fig. 4b (compare to the noiseless case in Fig. 4a).  While in Fig. 4a we see the \textit{signal} being spectrally reshaped due the HFT at frequency multiple or submultiple of $f_0$, in Fig. 4b we observe the \textit{noise} being similarly affected by the HFT, i.e. reshaped from the original flat spectrum to forming peaks at the eigen-frequency $f_0$ (see more details for noise shaping effect, also in \cite{ANS}). These two effects  (signal and noise shaping) being in resonance  enhance an amplitude of spiking (compare spiking amplitudes in Figs. 4a, 4b).  

Finally, we consider the same noisy input, but assume the HFT operating at the same eigen-frequency $f_0=0.2$ as a symbol frequency $f_s=0.2$. Note in Fig. 4c  a significant enhancement of spectral peaks and the corresponding amplification of spikes. 


\subsection{Flexibility of the HFT - multi-Harmonic fractal transformation}
Here we have shown that using the HFT one may model complex effects occurring in the FHN and characteristic to biological neurons. We have made the HFT as simple as possible, highlighting the frequency interplay wrapped around the frequency-amplitude dependency. 

Note, that above for simplicity we used only a main harmonic of the HFT. Similarly, one can extend the aforementioned analysis to multiple harmonics by optimizing weights to model spikes of any pulse-shaping as desired. This is similar to the Fourier transformation and series. 

As an example, we consider a case of replicating the main harmonics at odd integer frequency, i.e. $(2n+1)f_0$ with the weights $W_n=(2n+1)^{-1}$ , the results are plotted in Fig. 5 for same the inputs and parameters as in Fig. 2. Straightforwardly, the case in Fig. 2a from a simple sine-wave is transformed to a more complex shape in Fig. 5a by adding more harmonics as described. Various types of bursts and their amplitude dependency on input can  be achieved, see Fig. 5b, while for below-the-threshold cases a constant input may be fully suppressed (see Fig. 5c), while periodic inputs are almost completely preserved (see Fig. 5d).

\section*{Conclusions}
We have demonstrated that the HFT may recreate a vast range of complex effects characteristic and specific to spiking neuronal models. Unlike conventional filtering, the HFT does not include multiplication with the filter weights, instead performs summation of the signal spectra components with a frequency-dependent number of terms. Thus, it acts as an attractor in the spectral domain, modeling the noise shaping and spiking, while the fractal nature of the transformation enables to model frequency-dependent amplitude reshaping. Overall, the HFT models a vast variety of pulse shapes and  effects, as well as, provides a flexibility for optimization, while representing a simple frequency transformation at its core.

\end{document}